# Impact of Traffic Conditions and Carpool Lane Availability on Peer-to-Peer Ridesharing Demand

Sara Masoud and Young-Jun Son
Department of Systems and Industrial Engineering
University of Arizona
Tucson, AZ, USA

Neda Masoud and Jay Jayakrishnan
Civil and Environmental Engineering
University of California Irvine
Irvine, CA, USA

## Abstract

A peer-to-peer ridesharing system connects drivers who are using their personal vehicles to conduct their daily activities with passengers who are looking for rides. A well-designed and properly implemented ridesharing system can bring about social benefits, such as alleviating congestion and its adverse environmental impacts, as well as personal benefits in terms of shorter travel times and/or financial savings for the individuals involved. In this paper, the goal is to study the impact of availability of carpool lanes and traffic conditions on ridesharing demand using an agent-based simulation model. Agents will be given the option to use their personal vehicles, or participate in a ridesharing system. An exact many-to-many ride-matching algorithm, where each driver can pick-up and drop-off multiple passengers and each passenger can complete his/her trip by transferring between multiple vehicle, is used to match drivers with passengers. The proposed approach is implemented in AnyLogic® ABS software with a real travel data set of Los Angeles, California. The results of this research will shed light on the types of urban settings that will be more recipient towards ridesharing services.

## 1. Introduction

Traffic congestion is a major problem in urban areas, which may easily lead to major economic loss every year [1]. In the U.S., the increasing population growth of urban areas has led to rise of demands for highway travels, which in return has strained the transportation infrastructure. Travel demands and physical highway features along with traffic influenced events are identified as three major causes of traffic congestion by the Federal Highway Administration [2]. The 2015 Urban Mobility Scorecard report [3] reveals that traffic congestion in 2014 in the U.S. caused a total delay of 6.9 billion hours and a waste of 3.1 billion gallons in fuel, which led to total loss of $160 billion 2014 dollars. In addition to the waste of fuel, the additional green gas emission of congestion contributes to climate change. American Highway User Alliance has identified the top 30 highway bottlenecks in the U.S. [4]. 10 of these bottlenecks which are responsible for about 91 million hours of delay every year are located in the Los Angeles County, and have led to a queue length range of one to seven miles [4].

Various approaches have been implemented in the past to eliminate or to mitigate traffic congestion. One of the common approaches is expanding the infrastructure by adding new lanes to existing highways or even building new roads, which is costly. Texas A&M Transportation Institute showed that adding new lanes to an existing highway may cost about 2 to 10 million dollars per lane-mile, and the construction cost of a new highway is in range of 5 to 20 million dollars per lane-mile [5]. The disadvantage of this approach is that not only is it expensive, but also extension of most major highways will induce higher levels of demand for the improved roads, not alleviating the congestion, and even encouraging people to have additional trips, which eventually lead to a potential growth in traffic demand [6]. To find the optimal location for the new road segments, network design policy is usually utilized in a discrete form to define the additional road segments with respect to budget restriction to minimize the total travelling cost of the network. Similarly, to find the potential road segments for additional lane, network design is applied in a continuous form to achieve the optimal capacity expansion of the highway network [7]. On the other hand, instead of expanding the existing infrastructure, congestion management can be done by utilizing the unused capacity of the currently existing highway. To this end, one of the popular approaches in this area which has been employed for decades is toll pricing. Due to the improvement of the technology and implementation of the intelligent transportation systems in the transportation network, the emergence of the real-time traffic data has led to appearance of new concepts such as Dynamic toll pricing for a better control over traffic flow [8]. The only drawback of this approach is its limitation to tollways and toll lanes [9].

Public transit (e.g., bus) is another approach that can successfully reduce the vehicle miles travelled by carrying a high number of passengers by one vehicle. The disadvantage of this approach is the lack of flexibility due to fixed schedules and routes. Complementary to public transit, other services such as para-transit (e.g., shuttles) or taxies are defined based on a similar concept to reduce to vehicle miles travelled. However, Para-transit systems are limited due to the fact that they are usually dependent on the federal funding [10]. In addition, taxis usually contributes to traffic congestion in large cities [11].

Although many of the mentioned alternatives have been employed to manage traffic congestion, none of them has been effective enough to eliminate or alleviate the traffic congestion. This motivates the researchers to look for better alternatives such as shared-use mobility [12]. In the past few years, ridesharing has evolved from an idea into a reality. Ridesharing is the sharing of the vehicles by passengers who have similar routes to reduce the vehicle miles travelled and traffic congestion. The trip can be managed by one of the passengers, which is known as informal ridesharing (e.g., the carpooling for school or work by parents or collages) or it can be monitored by an external transportation network company (e.g., Uber or Lyft) to maximize the profit for both the company and the driver. As a result, the latter case can lead to additional traffic congestion like the case of taxis instead of solving the problem. In this paper, we consider alleviating traffic congestion and its adverse impacts, by using a peer-to-peer ridesharing system. In this system, drivers try to share their vehicle by other passengers who have similar routes while doing their daily activities. In addition, we will study the impact of availability of carpool lanes, and traffic conditions on ridesharing demand. To this end, a methodology that consists of agent-based simulation and a matching algorithm is introduced in Section 2. The experiment is discussed in Section 3. And, Section 4 provides the conclusions of the research works.

## 2. Methodology

The core of each ridesharing system is the ride-matching algorithm which can be considered as a general case of dial-a-ride problem [13]. In a practical dial-a-ride problem, several users request a door to door transportation service from a specific agency. Users can specify a time window for them to be picked up at the origin or be dropped off at the destination, or both. Agencies usually provide multiple vehicles to serve their customers. Each vehicle can handle multiple trip request in an each mission. Although an agency can minimize its operational costs by increasing the occupancy of its fleet, such savings come at the cost of low quality of service for users (due to longer trips) and may impact demand in the future. Drivers in a dial-a-ride system are employed by the company. In addition, it is assumed that all vehicles are available throughout the day. These two properties of a dial-a-ride system is what differentiates it from a peer-to-peer ridesharing system. In a ridesharing system, drivers are available only for a short period of time and within a certain time window due to their daily schedules, and each has a specific origin and destination of their own.

The limited availability of drivers in a peer-to-peer ridesharing problem make it really hard to achieve high matching rates in such systems [14]. To this end, Masoud and Jayakrishnan [10] have proposed an advanced ride-matching algorithm that can increase the efficiency of a ridesharing system by maximizing the number of successful driver-rider matches. Although the main difference between a dial-a-ride and a peer-to-peer ridesharing system is the fact that in the latter there is no pre-deterministic set of resources (i.e., vehicles), Masoud and Jayakrishnan showed that following and revising a basic dial-a-ride model may not be the most effective way to model a peer-to-peer ridesharing system [10]. They proposed a peer-to-peer ride-matching algorithm that can provide multi-hop routes for riders by optimally routing drivers in real-time. One of the properties of this algorithm is that it allows for using time-dependent travel times. This property becomes increasingly important in this paper, since the objective of this study is to measure the impact of carpool lanes on the percentage of successful matches, and in this context the attractiveness of carpooling is mainly in lower travel times. In this study, in order to take advantage of this property of the proposed system, we use agent-based simulation to incorporate realistic traffic conditions and travel times.

Here, the simulation model maintains the traffic equilibrium under different conditions and will call the matching algorithm whenever a potential rider enters the system. We assume that real-time travel times at both ordinary and carpool lanes are available to all the drivers at all times. Additionally, the model considers the network elements that may affect the interaction of vehicles such as the network topology, and speed limits. More details about the simulation model and algorithms are provided in the following sections.

**2.1. Agent-Based Simulation Model**

The agent-based model in this study provides a realistic environment, which allows for evaluating the efficiency of peer-to-peer ridesharing system. Although a ride-matching algorithm plays an important role in the effectiveness of the system, the successful implementation of a ridesharing system also depends on other elements, such as traffic conditions. The model has been constructed in AnyLogic® by utilizing the car following and lane changing libraries that are provided by the software [15]. In this model, three types of agents are defined. An agent can be a regular driver who is not participating in the peer-to-peer ridesharing experiment, a rider who is looking to employ a driver who can match his requirements, or a driver who volunteers to share his vehicle by other riders as long as he can follow his personal schedule. A rider can turn to a regular driver if he/she is not matched with any other driver in the ridesharing system. In this model, the regular drivers have their own individual behaviour and goal which is minimizing the total travelling cost. Their behaviour is modelled by Dijkstra's algorithm [16]. Each time a regular driver arrives to a decision making point (e.g., a highway junction), the driver will get the up-to-date traffic data of the network and by utilizing the Dijkstra's algorithm updates his/her route plan to his/her final destination. The following equation represents the traveling cost for each driver.

$$c_{ij}(t) = w_1(T_{ij}(t)) + w_2 s(f_{ij}(t)) \quad (1)$$

This equation shows that the travelling cost for each link at any moment in time ($C_{ij}(t)$) depends on average travelling time of the link (i.e., $s(f_{ij}(t))$) and also the potential toll price of the link (i.e. $T_{ij}(t)$) if the user decides to utilize a toll road. We denote the average travelling time as $s(f_{ij}(t))$ to emphasize that the traveling time depends on the traffic flow (i.e., $f_{ij}(t)$) of the link. This value is directly obtained from the simulation model. $W_1$ and $W_2$ are the coefficients to normalize the effect of toll price and average travelling time on the travelling cost.

Table 1: Validation of the simulation model

| Link ID | Traffic Flow Distribution | | Error |
| --- | --- | --- | --- |
| | Real Data | Simulation | |
| 0 | 0.056 | 0.054 | 0.002 |
| 1 | 0.231 | 0.227 | 0.004 |
| 2 | 0.460 | 0.464 | 0.005 |
| 3 | 0.253 | 0.254 | 0.001 |
| Average | | | 0.003 |

Table 1 compares the traffic distribution of the real data shown in Table 2 and the distribution provided by the simulation model. The last column, which shows the errors (i.e., absolute difference) between the real and simulated traffic flows, has an average of 0.003. Utilizing the data in Table 1 for a chi-squared test at 0.05 level of significance reveals that the null hypothesis (i.e., H0: The agent-based simulation data follow the real traffic flow distribution) cannot be rejected, and the simulated traffic flows can be considered as a realistic representation of the traffic flows of the area under study.

On the other hand, the main goal for the riders and drivers who participate in the ridesharing system is to minimize the travelling time while utilizing others' vehicles and maximizing the profit by sharing their vehicle by other riders, respectively. To model their behaviours, the matching algorithm would be utilized each time a rider enters the system. The next section provides details about the matching algorithm.

### 2.2. Matching Algorithm

In this work, a simulation model is used to incorporate the dynamic behaviour of traffic flow. As a result, a dynamic algorithm is needed, where users are allowed to register in the system when they are ready to travel. The ride-matching algorithm proposed by [10] is adopted in this work. The ride-matching algorithm not only creates a best set of driver-riders match in real-time, but also guarantees the optimal route plans for the participants. The steps used by the algorithm to find the optimal solution is depicted in Figure 1.

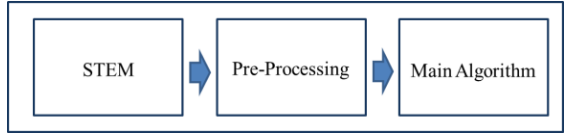

Figure 1: Logical steps in the matching algorithm

The algorithm consists of three main steps. In the first step, by using Spatio-Temporal Ellipsoid Method (STEM), a time-expanded network for each individual rider is generated, which is used to find the optimal route plan for the mentioned rider. To do so, the links whose starting and/or ending nodes are not reachable by the drivers in the original network are removed. In the remaining network, STEM finds time interval during which nodes are available based on the travel time window of the rider. Then, considering the rider's time intervals, STEM defines the feasible network for the rider.

The second step in the algorithm, as it is denoted in Figure 1, is pre-processing. Here, all the nodes that are not descendants of the origin and predecessors of the destination nodes are removed using an iterative process. Then, this revised network is topologically sorted. Finally, a depth fist search is employed to ensure that there exists at least one feasible route plan for the rider. If so, a dynamic programming algorithm is used to search the graph for the optimum route plan. The optimum route for each rider who enters the node $j$ on driver $d$'s vehicle is the one which satisfies the following Bellman equation:

$$V(n_j, d) = \min_{n_i: (n_i, d) \in DN_j} \left\{ \min_{d' \in D_i^{in} \setminus ED(n_i, d')} \{V(n_j, d') + C(n_i, n_j)\} \right\} \quad (2)$$

In Equation (2), $V(n_j,d)$ is the optimum cost of the optimal route from node 1 to node $j$. $DN_j$ is the set of all nodes $i$ engering node $j$ with $d$ as the driver.. $D_i^{in}$ is the set that contains all the drivers who enter node $i$. $ED(n_i, d')$ contains the list of optimal drivers on the optimal route to $n_i$, excluding the drivers in the last link to ensure the feasibility of the solution. $C(n_i, n_j)$ is the cost of traveling of the link connecting the two adjacent nodes $n_i$ and $n_j$. This cost is given by the simulation model if there is travelling between physically different nodes on the revised feasible time-expanded network, and can be equal to a penalty $P$, if the movement is only temporal. After running the matching algorithm, each driver knows where to pick up his/her assigned riders and where to drop them off. In addition, each rider will know if he/she can utilize the ridesharing service, and if so, when he will be picked up in a few seconds.

## 3. Experiments and Results

As mentioned in Section 1, the main goal of this study is to understand the effect of carpool lanes and traffic conditions on a dynamic peer-to-peer ridesharing system. To this end, we used a data set of Los Angeles County in Los Angeles, California, to construct a simulation model. The constructed simulation model is then used to conduct experiments to study the impacts of traffic conditions and carpool lanes. As mentioned previously, Los Angeles hosts eleven of the top thirty bottlenecks of the U.S. and out of these eleven bottlenecks, ten of them are in Los

Angeles County [4]. Figure 2 and Table 2 depict the part of the Los Angeles County that was modeled in AnyLogic® software.

Table 2: Topology of the testbed (i.e., part of Los Angeles County, California, 2014)

| Link | | | Length (mile) | Free Flow Travel Time (hour) | Carpool Lane | Traffic Flow (car / 24 hours) |
|---|---|---|---|---|---|---|
| ID | Starting Node | Ending Node | | | | |
| 0 | 0 | 3 | 36.0 | 0.55 | No | 12,948 |
| 1 | 0 | 1 | 14.3 | 0.22 | No | 53,319 |
| 2 | 1 | 2 | 33.1 | 0.50 | Yes | 106,058 |
| 3 | 1 | 3 | 41.7 | 0.64 | No | 58,343 |
| | Total | | 125.1 | 1.91 | NA | 230,668 |

As shown in Table 2, each link is defined by a unique ID. For each link (i.e., each row of the table), we gathered various data such as the length, the free flow travel time, the existence of a carpool lane, and the traffic flow of each link [17]. These pieces of information were used to validate the simulation model as well.

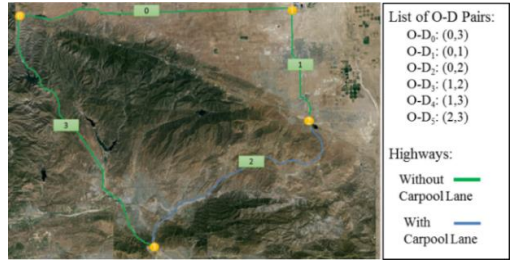

Figure 2: Highway topology of the testbed

Figure 2 depicts the topology of the network of our testbed which consist of four links. Out of these four links, only one of them (i.e., link 2 in blue) has a carpool lane and the other three green links do not have any carpool lanes. We considered six different Origin-Destination (O-D) pairs in this study, and all of them are shown in Figure 2. In this work, we conducted four different experiments to study the impact of carpool lanes and traffic flow on ridesharing. Table 3 provides the settings of each experiment as well as the results.

Table 3: Experiments and results

| Experiment ID | Unused Capacity of carpool lane | Participants | | | Successful Match rate |
|---|---|---|---|---|---|
| | | Rider | Driver | Regular driver | |
| 1 | 100% | 10% | 40% | 50% | 60% |
| 2 | 75% | 10% | 40% | 50% | 60% |
| 3 | 50% | 10% | 40% | 50% | 50% |
| 4 | 25% | 10% | 40% | 50% | 45% |

In the conducted experiments, the behavior and preference of the riders (e.g., the earliest arrival, latest arrival, earliest departure and latest departure), and the percentage of participants were fixed. In each experiment, different amount of the capacity of the carpool lane (i.e., column two) was used. The minimal percentage of unused capacity of carpool lane was 25%, because this percentage leads to an equivalent traffic flow in the carpool lane and other lanes. Table 3 shows that remaining capacity of carpool lane and the success rate in matching of the riders and drivers in a peer-to-peer ridesharing system has a direct relationship. It means that the lower traffic flow in the carpool lane leads to a higher rate in successfully matching the riders and drivers, which will reduce the traffic flow. In addition, it is important to note that other elements such as the time windows that the riders define or the topology of the system have significant impacts on the results. Smaller time windows make the matching harder. If the time intervals that the rider defines are smaller than a threshold the matching will be infeasible. The mentioned threshold dynamically dependent on the traffic flow of the related routes. The first experiment revealed that even with a practically empty carpool lane, only 60% of the riders could find a matching driver. On the other hand, the last experiment revealed that with a zero capacity carpool lane, still 45% of the riders can find their matching drivers in the system.

## 4. Conclusion

In this study, we proposed a simulation based approach involving a peer-to-peer ridesharing algorithm to study the impact of traffic conditions and carpool lane on the peer-to-peer ridesharing demand. This study was conducted based on a data set of Los Angeles County. An agent-based simulation was employed to present a realistic traffic setting by considering the impacts of the topology of the network and dynamic behavior of vehicles (e.g., car following and lane changing). Using the constructed simulation, four experiments were conducted to study the impacts of carpool lane and traffic conditions. The experimental results revealed a positive relation between carpool lane and a peer-to-peer ridesharing system.